%% file: make_astro.tex
\documentclass{mpe_report}

\usepackage{psfig,graphicx,epsfig}
\usepackage{color}
\usepackage{amsmath,amssymb,epic,eepic,array}

\begin{document}

\pagenumbering{arabic}
\setcounter{page}{165}

 \renewcommand{\FirstPageOfPaper }{165}\renewcommand{\LastPageOfPaper }{168}\include{./mpe_report_ikhsanov}        \clearpage

\end{document}

%% file: mpe_report_ikhsanov.tex
\newcommand{\dmfc}{\dot{\mathfrak{M}_{\rm c}}}
\newcommand{\dmfa}{\dot{\mathfrak{M}_{\rm a}}}
\newcommand{\dmf}{\dot{\mathfrak{M}}}
\newcommand{\be}{\begin{equation}}
\newcommand{\ee}{\end{equation}}
\newcommand{\bdm}{\begin{displaymath}}
\newcommand{\edm}{\end{displaymath}}

\title{Accreting Isolated Neutron Stars}
\author{N.R.\,Ikhsanov\inst{1,3} \and P.L.\,Biermann\inst{2,4}}
\institute{Institute of Astronomy, University of Cambridge,
           Madingley Road, Cambridge CB3\,0HA, UK
             \and
           Max-Planck-Institut f\"ur Radioastronomie, Auf dem
              H\"ugel 69, D-53121 Bonn, Germany
              \and
              Central Astronomical Observatory of the Russian
              Academy of Sciences, Pulkovo 65/1, 196140
              St.\,Petersburg, Russia
              \and
              Department of Physics and Astronomy, University of Bonn,
              Germany}
\maketitle

\begin{abstract}
Accretion of interstellar material by a magnetized, slowly rotating isolated neutron star
is discussed. We show that the average persistent X-ray luminosity of these objects is
unlikely to exceed $4 \times 10^{26}\,{\rm erg\,s^{-1}}$. They can also appear as X-ray
bursters with the burst duration of $\sim 30$\,minutes and repetition time of $\sim
10^5$\,yr. This indicates that the number of the accreting isolated neutron stars which
could be observed with recent X-ray missions is a few orders of magnitude smaller than
that previously estimated. Our findings argue against models in which the magnetic field
of neutron stars is assumed to decay exponentially on a time scale shorter than 500\,Myr.
\end{abstract}

\section{Introduction}

As shown first by Ostriker et~al. (\cite{Ostriker-etal-1970}) and Shvartsman
(\cite{Shvartsman-1971}) an old isolated neutron star moving through the interstellar
medium is able to capture material with a rate of \be\label{dmf0} \dmfc \simeq 10^9\,{\rm
g\,s^{-1}}\ \times\ n\ m^2\ \left(\frac{V_{\rm rel}}{10^7\,{\rm cm\,s^{-1}}}\right)^{-3}.
\ee Here $m$ is the mass of the neutron star expressed in units of $1.4\,M_{\sun}$, $n$
is the number density of the interstellar material expressed in units of 1 hydrogen atom
cm$^{-3}$ and $V_{\rm rel}$ is the relative velocity between the star and its
environment, which is limited to the sound speed in the interstellar material as $V_{\rm
rel} > V_{\rm s} \simeq 10^6\,{\rm cm\,s^{-1}}$. If all the captured material were
accreted onto the stellar surface the star would appear as an X-ray source of a
luminosity \be\label{l0} L_{\rm x} \la L_0 \simeq 10^{29}\,{\rm erg\,s^{-1}}\ \times\ m\
r_6^{-1}\ (\dmfa/\dmfc), \ee where $r_6$ is the radius of the neutron star expressed in
units of $10^6$\,cm and the parameter $\dmfa$ denotes the mass accretion rate onto the
stellar surface, which in the general case can differ from the mass capture rate by the
star from its environment, $\dmfc$. Combining this finding with currently established
spatial and velocity distributions of neutron stars one could expect about $3 \times
10^4$ such sources to be detected by {\it Chandra} and XMM-{\it Newton} (for a discussion
see, e.g., Popov et~al. \cite{Popov-etal-2000a}). However, none of them has been
identified so far.

A lack of success in searching for the accreting isolated neutron stars indicates that
either the value of $\dmfc$ is significantly smaller than that evaluated from
Eq.~(\ref{dmf0}) or there is an additional factor which prevents the interstellar
material from reaching the stellar surface. The first possibility has been critically
discussed and basically discarded by Perna et~al. (\cite{Perna-etal-2003}). Among factors
which have not been taken into account in the above accretion scenario is the magnetic
field of the neutron star. A zero-field approximation appears to be reasonable only if
the dipole magnetic moment of an isolated neutron star is $< 10^{22}\,{\rm G\,cm^3}$ (the
corresponding surface field is $\sim 10^4$\,G). Under this condition a formation of the
magnetosphere which could prevent the captured material from reaching the stellar surface
does not occur. If, however, the surface field significantly exceeds 10\,kG the influence
of the stellar magnetic field on the accretion picture has to be taken into account. As
we show the expected X-ray luminosity of accreting isolated neutron stars in this case
proves to be significantly smaller than that estimated by Eq.~(\ref{l0}).

  \section{Magnetic field strength and spin period}

A necessary condition for the captured material to reach the surface of a magnetized,
neutron star rotating with a period $P_{\rm s}$ is \be\label{main} r_{\rm m} < r_{\rm
cor}, \ee where \be r_{\rm m} \simeq 6\times 10^{10}\,{\rm cm}\ \times\ \mu_{30}^{4/7}\
\dmf_9^{-2/7}\ m^{-1/7}, \ee is the magnetospheric radius of a neutron star, and \be
r_{\rm cor} \simeq 1.7 \times 10^8\,{\rm cm}\ \times\  m^{1/3}\ P_{\rm s}^{2/3} \ee is
its corotational radius. Here $\mu_{30}$ is the dipole magnetic moment of the star
expressed in units of $10^{30}\,{\rm G\,cm^3}$ and $\dmf_9=\dmfc/10^9\,{\rm g\,s^{-1}}$.
Solving the inequality~(\ref{main}) for $P_{\rm s}$ one finds \be P_{\rm s} > P_{\rm cd}
\simeq 7000\ {\rm s}\ \times\ \mu_{30}^{6/7} V_7^{9/7} n^{-3/7} m^{-11/7}, \ee where $V_7
= V_{\rm rel}/10^7\,{\rm cm\,s^{-1}}$. This implies that the spin-down rate of the
neutron star in a previous epoch was \be \dot{P} > 10^{-14}\ \left[\frac{P_{\rm
s}}{7000\,{\rm s}}\right] \left[\frac{t_{\rm sd}}{10^{10}\,{\rm yr}}\right]^{-1}\ {\rm
s\,s^{-1}},
   \ee
and therefore, suggests that only the stars with initial dipole magnetic moment in excess
of $10^{29}\,{\rm G\,cm^3}$ could be a subject of further consideration (for a discussion
see, e.g., Popov et~al. \cite{Popov-etal-2000b}). Here $t_{\rm sd}$ is the spin-down
timescale of the neutron star.

  \section{Accretion flow geometry}

For an accretion disk around a magnetized isolated neutron star to form the relative
velocity should satisfy the condition $V_{\rm rel} < V_0$, where
  \be
V_0 \simeq 10^5\ {\rm cm\,s^{-1}}\ \times\ \mu_{30}^{-6/65} n^{3/65} m^{5/13}\ \times \ee
\bdm \left(\frac{V_{\rm t}}{10^6\,{\rm cm\,s^{-1}}}\right)^{21/65} \left(\frac{R_{\rm
t}}{10^{20}\,{\rm cm}}\right)^{-7/65}. \edm Here $V_{\rm t}$ is the velocity of turbulent
motions of the interstellar material at a scale of $R_{\rm t}$ and the Kolmogorov
spectrum of the turbulent motions is assumed (Prokhorov et~al.
\cite{Prokhorov-etal-2002}). This inequality, however, is unlikely to be satisfied since
$V_0$ is smaller than the speed of sound in the interstellar material, and therefore, is
smaller than the lower limit to $V_{\rm rel}$.

Thus, the accretion by old isolated neutron stars can be treated in terms of a spherical
(Bondi) accretion onto a magnetized, slowly rotating neutron star. The accretion picture
under these conditions has been reconstructed first by Arons \& Lea
(\cite{Arons-Lea-1976}) and Elsner \& Lamb (\cite{Elsner-Lamb-1976}) and further
developed by Lamb et~al. (\cite{Lamb-etal-1977}) and Elsner \& Lamb
(\cite{Elsner-Lamb-1984}). An application of the results reported in these papers to the
case of an isolated neutron star is discussed in the following section.

   \section{Accretion flow at $r_{\rm m}$}

As shown by Arons \& Lea (\cite{Arons-Lea-1976}) and Elsner \& Lamb
(\cite{Elsner-Lamb-1976}), the magnetosphere of a neutron star undergoing spherically
symmetrical accretion is closed and, in the first approximation, prevents the accretion
flow from reaching the stellar surface. The mass accretion rate onto the stellar surface
is therefore limited to the rate of plasma entry into the magnetosphere. The fastest
modes by which the material stored over the magnetospheric boundary can enter the stellar
field are the Bohm diffusion and interchange instabilities (Elsner \& Lamb
\cite{Elsner-Lamb-1984}).

The rate of plasma diffusion in the case considered can be evaluated as (Ikhsanov
\cite{Ikhsanov-2003}) \be \dmf_{\rm B} \leq 2 \times 10^6\,{\rm g\,s^{-1}}\
\zeta_{0.1}^{1/2}\ \mu_{30}^{-1/14}\ m^{15/7}\ n^{11/14}\ V_7^{33/14},
  \ee
where $\zeta_{0.1} = \zeta/0.1$ is the efficiency of the diffusion process normalized
according to Gosling et~al. (\cite{Gosling-etal-1991}). This indicates that the
luminosity of the diffusion-driven source is limited to \be\label{lxdd} L_{\rm x,dd} \leq
4 \times 10^{26}\,{\rm erg\,s^{-1}}\ \times \ee \bdm \zeta_{0.1}^{1/2}\ \mu_{30}^{-1/14}\
m^{22/7}\ n^{11/14}\ V_7^{33/14}\ r_6^{-1}. \edm

For the material to enter the stellar magnetic field with the rate $\sim \dmfc$ the
boundary should be interchange unstable. The onset condition for the instabilities is
(Arons \& Lea \cite{Arons-Lea-1976}; Elsner \& Lamb \cite{Elsner-Lamb-1976})
\be\label{inst} T_{\rm p}(r_{\rm m}) \leq 0.1 T_{\rm ff}(r_{\rm m}), \ee where $T_{\rm
p}(r_{\rm m})$ and $T_{\rm ff}(r_{\rm m})$ are the plasma temperature and the free-fall
(adiabatic) temperature at the magnetospheric boundary, respectively. This indicates that
a direct accretion of the captured material onto the stellar surface could occur only if
the cooling of the plasma at the boundary dominates the heating.

The mechanism which is responsible for the cooling of plasma at the boundary is the
bremsstrahlung emission. Indeed, the free-fall temperature and number density of the
material stored over the boundary are, respectively, \be T_{\rm ff}(r_{\rm m}) \simeq\
10^7\,{\rm K}\ \times\ \mu_{30}^{-4/7} \dmf_9^{2/7} m^{6/7}, \ee \be N_{\rm e}(r_{\rm m})
\simeq 300\,{\rm cm^{-3}}\ \times\ \mu_{30}^{-6/7} \dmf_9^{10/7} m^{-2/7}. \ee Under
these conditions both the cyclotron and Compton cooling time scale are significantly
larger than the bremsstrahlung cooling time scale \be t_{\rm br}(r_{\rm m}) \simeq
10^5\,{\rm yr}\ \times\ T_7^{1/2} \left(\frac{N_{\rm e}(r_{\rm m})}{300\,{\rm
cm^{-3}}}\right)^{-1}, \ee where $T_7 = T_{\rm ff}(r_{\rm m})/10^7$\,K.

The heating of the material at the magnetospheric boundary is governed by the following
processes.

  \subsection{Adiabatic shock}\label{shock}

As the captured material reaches the boundary it stops in an adiabatic shock. The
temperature in the shock increases to $T_{\rm ff}(r_{\rm m})$ on a dynamical time scale,
\be t_{\rm ff}(r_{\rm m}) \simeq 740\,{\rm s}\ \times\ m^{-1/2} \left(\frac{r_{\rm
m}}{6\times 10^{10}\,{\rm cm}}\right)^{3/2}. \ee Since $t_{\rm ff}(r_{\rm m}) \ll t_{\rm
br}(r_{\rm m})$ the height of the homogeneous atmosphere at the boundary proves to be
$\sim r_{\rm m}$. This prevents an accumulation of material over the boundary.
Furthermore, as the condition $t_{\rm ff}(r_{\rm G}) < t_{\rm br}(r_{\rm m})$ is
satisfied throughout the gravitational radius of the neutron star a hot quasi-stationary
envelope extended from $r_{\rm m}$ up to $r_{\rm G}$ forms (Davies \& Pringle
\cite{Davies-Pringle-1981}). The formation of the envelope prevents the surrounding
material from pene\-trating to within the gravitational radius of the neutron star. The
mass of the envelope is, therefore, conserved. As the neutron star moves through the
interstellar medium the surrounding material overflow the outer edge of the envelope
with a rate of $\dmfc$.

Within an approximation of a non-rotating star whose ``magnetic gate'' at the boundary is
closed completely the envelope remains in a stationary state on a time scale of $t_{\rm
br}(r_{\rm m})$. As the condition~(\ref{inst}) is satisfied the boundary becomes unstable
and material enters into the magnetic field and accretes onto the stellar surface with a
rate of $\sim \dmfc$. As shown by Lamb et~al. (\cite{Lamb-etal-1977}), the time of the
accretion event in this case is limited to $t_{\rm burst} <$\,a few\,$\times t_{\rm
ff}(r_{\rm m})$ during which the temperature of the envelope increases again to the
adiabatic temperature (as the upper layers of the envelope come to $r_{\rm m}$). The
corresponding source, therefore, would appear as an X-ray burster with the luminosity \be
L_{\rm burst} \simeq 2 \times 10^{29}\ n\ V_7^{-3}\ m^3\ r_6^{-1}\ {\rm erg\,s^{-1}}, \ee
the typical outburst durations of $t_{\rm burst} \leq 30$\,min and the repetition time of
$t_{\rm rep} \sim t_{\rm br}(r_{\rm m}) \sim 10^5$\,yr.

  \subsection{Subsonic propeller}

As shown by Davies \& Pringle (\cite{Davies-Pringle-1981}), the rotation of a neutron
star surrounded by the hot envelope can be neglected only if its spin period exceeds
(Ikhsanov \cite{Ikhsanov-2001}) \be P_{\rm br} \simeq 10^5\,{\rm s}\ \times\
\mu_{30}^{16/21} n^{-5/7} V_7^{15/7} m^{-34/21}. \ee Otherwise, the heating of plasma at
the inner edge of the envelope due to propeller action by the star dominates cooling. The
corresponding state of the neutron star is referred to as a subsonic propeller. The star
remains in this state as long as its spin period satisfies the condition $P_{\rm cd} <
P_{\rm s} < P_{\rm br}$. The time during which the spin period increases from $P_{\rm
cd}$ up to $P_{\rm br}$ is \be \tau_{\rm br} \simeq 2 \times 10^5\,{\rm yr}\ \times\
\mu_{30}^{-2}\ I_{45}\ m\ \left(\frac{P_{\rm br}}{10^5\,{\rm yr}}\right), \ee where
$I_{45}$ is the moment of inertia of the neutron star expressed in units of
$10^{45}\,{\rm g\,cm^2}$. This indicates that the spin periods of accreting isolated
neutron stars are expected to be in excess of a day, and therefore, these objects are
unlikely to be recognized as pulsars.

  \subsection{Diffusion-driven accretor}

As mentioned above, the ``magnetic gate'' at the magnetospheric boundary is not closed
completely. The plasma flow through the interchange stable boundary is governed by the
diffusion. As shown by Ikhsanov (\cite{Ikhsanov-2003}), this leads to a drift of the
envelope material towards the star and, as a result, to an additional energy source for
heating of the envelope material. The heating due to the radial drift dominates the
bremsstrahlung energy losses from the envelope if $\dmfc < \dmf_{\rm cr}$, where \be
\dmf_{\rm cr} \simeq 10^{14}\,{\rm g\,s^{-1}}\ \times\ \zeta_{0.1}^{7/17}
\mu_{30}^{-1/17} V_{7}^{14/17} m^{16/17}. \ee This indicates that only the old isolated
neutron stars which move slowly ($V_{\rm rel} \ll 10^7\,{\rm cm\,s^{-1}}$) through a
dense molecular cloud ($N_{\rm e} > 10\,{\rm cm^{-3}}$) can be expected to be observed as
the bursters. The rest of the population would appear as persistent X-ray sources with
the luminosity of $L_{\rm x} \leq L_{\rm x,dd}$ (see Eq.~\ref{lxdd}).

  \section{Discussion}

The results of this paper force us to reconsider previous predictions about the number of
old isolated neutron stars which can be observed with current X-ray missions. In
particular, if the surface field strength of these objects exceeds 10\,kG the total flux
of their persistent X-ray emission is limited to $F < 10^{-16}\,{\rm
erg\,cm^2\,s^{-1}}\,d_{100}^{-2}$, where $d_{100}$ is the distance to the source
expressed in units of 100\,pc. The mean energy of the emitted photons within the
blackbody approximation is close to 50\,eV. This clearly shows that detection of these
sources by {\it Chandra} and XMM-{\it Newton} is impossible.

The X-ray flux emitted during the outbursts (see Sect.\,\ref{shock}) is over the
threshold of sensitivity of modern detectors. However, the probability to detect this
event appears to be negligibly small. Indeed, the number of these sources which could be
detected by current X-ray missions is \be N \leq 10^{-5} \left(\frac{N(0)}{3 \times
10^4}\right) \left(\frac{t_{\rm burst}}{30\,{\rm min}}\right) \left(\frac{t_{\rm
rep}}{10^5\,{\rm yr}}\right), \ee where $N(0)$ is the number of the sources which would
be observed if the influence of the stellar magnetic field on the accretion flow at
$r_{\rm m}$ were insignificant (Popov et~al. \cite{Popov-etal-2000a}).

Finally, if the magnetic field of an isolated neutron star were $\la 10$\,kG it would
appear as an accretion-powered X-ray source with a luminosity of $L_0$ (see
Eq.~\ref{l0}). Indeed, the state of the star under these conditions can be unambiguously
identified with an accretor independently of its initial period and magnetic field
strength. Furthermore, as shown by Bisnovatyi-Kogan \& Blinnikov
(\cite{Bisnovatyi-Kogan-Blinnikov-1980}) a heating of the accretion flow inside the Bondi
radius is unable to reduce the mass accretion rate onto the stellar surface
significantly. That is why, a lack of success in searching for the accreting isolated
neutron stars indicates that the magnetic field of old neutron stars does not vanish
completely, but remains at least over a level of 10\,kG. This allows us to discard a
situation in which the magnetic field of a neutron star dissipates exponentially on a
time scale shorter that 500\,Myr.

\begin{acknowledgements}
Nazar Ikhsanov acknowledge the support of the European Commission under the Marie Curie
Incoming Fellowship Program. Support for Peter L. Biermann is coming from the AUGER
membership and theory grant 05 CU 5PD 1/2 via DESY/BMBF.
\end{acknowledgements}

%% file: make_astro.bbl
\begin{thebibliography}{}


\bibitem[1976]{Arons-Lea-1976}
Arons, J., \& Lea, S.M. 1976, ApJ, 207, 914

\bibitem[1980]{Bisnovatyi-Kogan-Blinnikov-1980}
Bisnovatyi-Kogan, G.S., \& Blinnikov, S.I. 1980, MNRAS, 191, 711

\bibitem[1981]{Davies-Pringle-1981}
Davies, R.E., \& Pringle, J.E. 1981, MNRAS, 196, 209

\bibitem[1976]{Elsner-Lamb-1976}
Elsner, R.F., \& Lamb, F.K. 1976, Nature, 262, 356

\bibitem[1984]{Elsner-Lamb-1984}
Elsner, R.F., \& Lamb, F.K. 1984, ApJ, 278, 326

\bibitem[1991]{Gosling-etal-1991}
Gosling, J.T., Thomsen, M.F., Bame, S.J., et al. 1991, J. Geophys. Res., 96, 14097

\bibitem[2001]{Ikhsanov-2001}
Ikhsanov, N.R. 2001, A\&A, 368, L5

\bibitem[2003]{Ikhsanov-2003}
Ikhsanov, N.R. 2003, A\&A, 399, 1147

\bibitem[1977]{Lamb-etal-1977}
Lamb, F.K., Fabian, A.C., Pringle, J.E., \& Lamb, D.Q. 1977, ApJ, 217, 197

\bibitem[1970]{Ostriker-etal-1970}
Ostriker, J.P., Rees, M.J., \& Silk, J. 1970, Astrophys. Lett. Commun., 6, 179

\bibitem[2003]{Perna-etal-2003}
Perna, R., Narayan, R., Rybicki, G., Stella, L., \& Treves, A. 2003, ApJ, 594, 936

\bibitem[2000a]{Popov-etal-2000a}
Popov, S.B., Colpi, M., Prokhorov, M.E., et~al. 2000a, ApJ, 544, L53

\bibitem[2000b]{Popov-etal-2000b}
Popov, S.B., Colpi, M., Treves, A., et al. 2000b, ApJ, 530, 896

\bibitem[2002]{Prokhorov-etal-2002}
Prokhorov, M.E., Popov, S.B. \& Khoperskov, A.V. 2002, A\&A, 381, 1000

\bibitem[1971]{Shvartsman-1971}
Shvartsman, V.F. 1971, Sov. Astron., 14, 662
\end{thebibliography}
